\documentclass[italian,english]{article}
\usepackage[T1]{fontenc}
\usepackage[latin9]{inputenc}
\usepackage{color}
\usepackage{textcomp}
\usepackage{amstext}
\usepackage{amssymb}

\makeatletter

\newcommand{\lyxmathsym}[1]{\ifmmode\begingroup\def\b@ld{bold}
  \text{\ifx\math@version\b@ld\bfseries\fi#1}\endgroup\else#1\fi}

\makeatother

\usepackage{babel}
\begin{document}

\title{\textbf{Non-strictly black body spectrum from the tunnelling mechanism }}

\author{\textbf{Christian Corda}}

\maketitle
\begin{center}
Institute for Theoretical Physics and Advanced Mathematics Einstein-Galilei,
Via Santa Gonda 14, 59100 Prato, Italy 
\par\end{center}

\begin{center}
\textit{E-mail address:} \textcolor{blue}{cordac.galilei@gmail.com} 
\par\end{center}
\begin{abstract}
A modern and largely used approach to obtain Hawking radiation is
the tunnelling mechanism. However, in various papers in the literature,
the analysis concerned almost only to obtain the Hawking temperature
through a comparison of the probability of emission of an outgoing
particle with the Boltzmann factor. 

In a interesting and well written paper, Banerjee and Majhi improved
the approach, by explicitly finding a black body spectrum associated
with black holes. On the other hand, this result, which has been obtained
by using a reformulation of the tunnelling mechanism, is in contrast
which the remarkable result by Parikh and Wilczek, that, indeed, found
a probability of emission which is compatible with a non-strictly
thermal spectrum.

By using our recent introduction of an effective state for a black
hole, here we solve such a contradiction, through a slight modification
of the analysis by Banerjee and Majhi. The final result will be a
non-strictly black body spectrum from the tunnelling mechanism. 

We also show that, for an effective temperature, we can write the
corresponding effective metric by Hawking's periodicity arguments.

Potential important implications for the black hole information puzzle
are also discussed.
\end{abstract}
In recent years, the tunnelling mechanism has been an elegant and
largely used approach to obtain Hawking radiation \cite{key-1}, see
for example \cite{key-2}-\cite{key-6} and refs. within. Let us consider
an object which is classically stable. If it becomes unstable from
a quantum-mechanically point of view, one naturally suspects tunnelling.
The mechanism of particles creation by black holes \cite{key-1},
can be described as tunnelling arising from vacuum fluctuations near
the black hole's horizon \cite{key-2}-\cite{key-6}. If a virtual
particle pair is created just inside the horizon, the virtual particle
with positive energy can tunnel out. Then, it materializes outside
the black hole as a real particle. In the same way, if one considers
a virtual particle pair created just outside the horizon, the particle
with negative energy can tunnel inwards. In both of the situations,
the particle with negative energy is absorbed by the black hole. The
result will be that the mass of the black hole decreases and the particle
with positive energy propagates towards infinity. Thus, subsequent
emissions of quanta appear as Hawking radiation. A problem on such
an approach was that, in the cited and in other papers in the literature,
the analysis has been finalized almost only to obtain the Hawking
temperature through a comparison of the probability of emission of
an outgoing particle with the Boltzmann factor. The problem was apparently
solved in the interesting work \cite{key-7}, where, through a reformulation
of the tunnelling mechanism, a black body spectrum associated with
black holes has been found. In any case, this result is in contrast
which the remarkable result in \cite{key-2,key-3}, that, indeed,
found a probability of emission which is compatible with a non-strictly
thermal spectrum. The non precisely thermal character of the spectrum
has important implications for the black hole information puzzle as
arguments that information is lost during black hole's evaporation
partially rely on the assumption of strict thermal behavior of the
spectrum \cite{key-2,key-3,key-8,key-9}. In fact, by introducing
an \emph{effective state}, we recently interpreted black hole's quasi-normal
modes in terms of quantum levels by finding a natural connection between
Hawking radiation and quasi-normal modes \cite{key-8,key-9}. As for
large $n$ black holes result to be well defined quantum mechanical
systems, having ordered, discrete quantum spectra, the results in
\cite{key-8,key-9} look consistent with the unitarity of the underlying
quantum gravity theory and with the idea that information should come
out in black hole's evaporation. 

Here we show that the \emph{effective quantities} permit also to solve
the above cited contradiction, through a slight modification of the
analysis in \cite{key-7}. The final result will be a non-strictly
black body spectrum from the tunnelling mechanism. 

For the sake of simplicity, in this letter we refer to the Schwarzschild
black hole and we work with $G=c=k_{B}=\hbar=\frac{1}{4\pi\epsilon_{0}}=1$
(Planck units).

Let us consider a Schwarzschild black hole. The Schwarzschild line
element is (see \cite{key-11} for clarifying historical notes to
this notion) 
\begin{equation}
ds^{2}=-(1-\frac{2M}{r})dt^{2}+\frac{dr^{2}}{1-\frac{2M}{r}}+r^{2}(\sin^{2}\theta d\varphi^{2}+d\theta^{2}).\label{eq: Hilbert}
\end{equation}
The event horizon is defined by $r_{H}=2M$ \cite{key-7,key-10},
while $\frac{1}{4M}\:$ is the black hole's surface gravity. As we
want to discuss Hawking radiation like tunnelling, the radial trajectory
is relevant \cite{key-2,key-3,key-7}. The analysis in \cite{key-7}
permitted to write down the (normalized) physical states of the system
for bosons and fermions as \cite{key-7} 
\begin{equation}
\begin{array}{c}
|\Psi>_{boson}=\left(1-\exp\left(-8\pi M\omega\right)\right)^{\frac{1}{2}}\sum_{n}\exp\left(-4\pi nM\omega\right)|n_{out}^{(L)}>\otimes|n_{out}^{(R)}>\\
\\
|\Psi>_{fermion}=\left(1+\exp\left(-8\pi M\omega\right)\right)^{-\frac{1}{2}}\sum_{n}\exp\left(-4\pi nM\omega\right)|n_{out}^{(L)}>\otimes|n_{out}^{(R)}>.
\end{array}\label{eq: physical states}
\end{equation}

Hereafter we focus the analysis only on bosons. In fact, for fermions
the analysis is identical \cite{key-7}. The density matrix operator
of the system is \cite{key-7}

\begin{equation}
\begin{array}{c}
\hat{\rho}_{boson}\equiv\Psi>_{boson}<\Psi|_{boson}\\
\\
=\left(1-\exp\left(-8\pi M\omega\right)\right)\sum_{n,m}\exp\left[-4\pi(n+m)M\omega\right]|n_{out}^{(L)}>\otimes|n_{out}^{(R)}><m_{out}^{(R)}|\otimes<m_{out}^{(L)}|.
\end{array}\label{eq: matrice densita}
\end{equation}
If one traces out the ingoing modes, the density matrix for the outgoing
(right) modes reads \cite{key-7}
\begin{equation}
\hat{\rho}_{boson}^{(R)}=\left(1-\exp\left(-8\pi M\omega\right)\right)\sum_{n}\exp\left(-8\pi nM\omega\right)|n_{out}^{(R)}><n_{out}^{(R)}|.\label{eq: matrice densita destra}
\end{equation}
This implies that the average number of particles detected at infinity
is \cite{key-7}

\begin{equation}
<n>_{boson}=tr\left[\hat{n}\hat{\rho}_{boson}^{(R)}\right]=\frac{1}{\exp\left(8\pi M\omega\right)-1},\label{eq: traccia}
\end{equation}
where the trace has been taken over all the eigenstates and the final
result has been obtained through a bit of algebra, see \cite{key-7}
for details. The result of eq. (\ref{eq: traccia}) is the well known
Bose-Einstein distribution. A similar analysis works also for fermions
\cite{key-7}, and one easily gets the well known Fermi-Dirac distribution

\begin{equation}
<n>_{fermion}=\frac{1}{\exp\left(8\pi M\omega\right)+1},\label{eq: traccia 2}
\end{equation}
Both the distributions correspond to a black body spectrum with the
Hawking temperature \cite{key-1,key-7}

\begin{equation}
T_{H}\equiv\frac{1}{8\pi M}.\label{eq: Hawking temperature}
\end{equation}

The result in \cite{key-7}, that we shortly reviewed, is remarkable.
In fact, through a reformulation of the tunnelling mechanism, one
can found a black body spectrum associated with black holes which
is in perfect agreement with the famous original result by Hawking
\cite{key-1}. On the other hand, it is in contrast with another remarkable
result \cite{key-2,key-3}. In fact, the probability of emission connected
with the two distributions (21) and (22) is given by \cite{key-1,key-2,key-3}
\begin{equation}
\Gamma\sim\exp(-\frac{\omega}{T_{H}}).\label{eq: hawking probability}
\end{equation}
But in \cite{key-2,key-3} a remarkable correction, through an exact
calculation of the action for a tunnelling spherically symmetric particle,
has been found, yielding 
\begin{equation}
\Gamma\sim\exp[-\frac{\omega}{T_{H}}(1-\frac{\omega}{2M})].\label{eq: Parikh Correction}
\end{equation}
This important result, which is clearly in contrast with the result
in \cite{key-7}, enables a correction, the additional term $\frac{\omega}{2M}$
\cite{key-2,key-3}. The important difference is that the authors
of \cite{key-7} did not taken into due account the conservation of
the energy, which generates a dynamical instead of static geometry
of the black hole \cite{key-2,key-3}. In other words, the energy
conservation forces the black hole to contract during the process
of radiation \cite{key-2,key-3}. Therefore, the horizon recedes from
its original radius, and, at the end of the emission, the radius becomes
smaller \cite{key-2,key-3}. The consequence is that black holes do
not strictly emit like black bodies \cite{key-2,key-3}. 

It is important to recall that the tunnelling is a \emph{discrete}
instead of \emph{continuous} process \cite{key-8}. In fact, two different
\emph{countable} black hole's physical states must be considered,
the physical state before the emission of the particle and the physical
state after the emission of the particle \cite{key-8}. Thus, the
emission of the particle can be interpreted like a \emph{quantum}
\emph{transition} of frequency $\omega$ between the two discrete
states \cite{key-8}. In the language of the tunnelling mechanism,
a trajectory in imaginary or complex time joins two separated classical
turning points \cite{key-2,key-3}. Another important consequence
is that the radiation spectrum is also discrete \cite{key-8}. Let
us clarify this important issue in a better way. At a well fixed Hawking
temperature and the statistical probability distribution (\ref{eq: Parikh Correction})
are continuous functions. On the other hand, the Hawking temperature
in (\ref{eq: Parikh Correction}) varies in time with a character
which is \emph{discrete}. In fact, the forbidden region traversed
by the emitting particle has a \emph{finite} size \cite{key-3}. Considering
a strictly thermal approximation, the turning points have zero separation.
Therefore, it is not clear what joining trajectory has to be considered
because there is not barrier \cite{key-3}. The problem is solved
if we argue that the forbidden finite region from $r_{initial}=2M\:$
to $r_{final}=2(M\lyxmathsym{\textminus}\omega)\:$ that the tunnelling
particle traverses works like barrier \cite{key-3}. Thus, the intriguing
explanation is that it is the particle itself which generates a tunnel
through the horizon \cite{key-3}.

A good way to take into due account the dynamical geometry of the
black hole during the emission of the particle is to introduce the
black hole's \emph{effective state}. By introducing the \emph{effective
temperature }\cite{key-8,key-9} 
\begin{equation}
T_{E}(\omega)\equiv\frac{2M}{2M-\omega}T_{H}=\frac{1}{4\pi(2M-\omega)},\label{eq: Corda Temperature}
\end{equation}
one re-writes eq. (\ref{eq: Corda Temperature}) in a Boltzmann-like
form similar to the original probability found by Hawking 
\begin{equation}
\Gamma\sim\exp[-\beta_{E}(\omega)\omega]=\exp(-\frac{\omega}{T_{E}(\omega)}),\label{eq: Corda Probability}
\end{equation}

\noindent where $\exp[-\beta_{E}(\omega)\omega]$ is the \emph{effective
Boltzmann factor,} with \cite{key-8,key-9}

\noindent 
\begin{equation}
\beta_{E}(\omega)\equiv\frac{1}{T_{E}(\omega)}.\label{eq: beta E}
\end{equation}
Hence, the effective temperature replaces the Hawking temperature
in the equation of the probability of emission \cite{key-8,key-9}.
Let us discuss the physical interpretation. In various fields of science,
we can takes into account the deviation from the thermal spectrum
of an emitting body by introducing an effective temperature. It represents
the temperature of a black body that would emit the same total amount
of radiation\emph{. }We introduced the concept of effective temperature
in the black hole's physics in \cite{key-8,key-9}. $T_{E}(\omega)$
depends on the energy-frequency of the emitted radiation and the ratio
$\frac{T_{E}(\omega)}{T_{H}}=\frac{2M}{2M-\omega}$ represents the
deviation of the radiation spectrum of a black hole from the strictly
thermal feature \cite{key-8,key-9}.

The introduction of $T_{E}(\omega)$ permits the introduction of others
\emph{effective quantities}. In fact, let us consider the initial
mass of the black hole \emph{before} the emission, $M$, and the final
mass of the hole \emph{after} the emission, $M-\omega$ respectively
\cite{key-8,key-9}. The \emph{effective mass }and the \emph{effective
horizon} of the black hole \emph{during} its contraction, i.e. \emph{during}
the emission of the particle, are defined as \cite{key-8,key-9} 
\begin{equation}
M_{E}\equiv M-\frac{\omega}{2},\mbox{ }r_{E}\equiv2M_{E}.\label{eq: effective quantities}
\end{equation}

\noindent The above effective quantities are average quantities\emph{
}\cite{key-8,key-9}. \emph{$r_{E}$ }is the average of the initial
and final horizons and \emph{$M_{E}\:$ }is the average of the initial
and final masses \cite{key-8,key-9}. Therefore, \emph{$T_{E}\:$
}is the inverse of the average value of the inverses of the initial
and final Hawking temperatures (\emph{before} the emission $T_{H\mbox{ initial}}=\frac{1}{8\pi M}$,
\emph{after} the emission $T_{H\mbox{ final}}=\frac{1}{8\pi(M-\omega)}$
respectively) \cite{key-8,key-9}. Thus, the Hawking temperature \emph{has
a discrete character in time}. 

We stress that the introduction of the effective temperature does
not degrade the importance of the Hawking temperature. Indeed, as
the Hawking temperature changes with a discrete behavior in time,
it is not clear which value of such a temperature has to be associated
to the emission of the particle. Has one to consider the value of
the Hawking temperature \emph{before} the \emph{emission} or the value
of the Hawking temperature after the emission? The answer is that
one must consider an \emph{intermediate} value, the effective temperature,
which is the inverse of the average value of the inverses of the initial
and final Hawking temperatures. In a certain sense, it represents
the value of the Hawking temperature \emph{during} the emission. $T_{E}(\omega)$
takes into account the non-strictly thermal character of the radiation
spectrum and the non-strictly continuous character of subsequent emissions
of Hawking quanta.

Therefore, one can define two further effective quantities. The \emph{effective
Schwarzschild line element }is given by 
\begin{equation}
ds_{E}^{2}\equiv-(1-\frac{2M_{E}}{r})dt^{2}+\frac{dr^{2}}{1-\frac{2M_{E}}{r}}+r^{2}(\sin^{2}\theta d\varphi^{2}+d\theta^{2}),\label{eq: Hilbert effective}
\end{equation}
and, consequently, the \emph{effective} \emph{surface gravity} is
defined as\emph{ }$\frac{1}{4M_{E}}.$ Thus, the effective line element
(\ref{eq: Hilbert effective}) takes into account the \emph{dynamical}
geometry of the black hole during the emission of the particle. Clearly,
if one follows step by step the analysis in \cite{key-7}, at the
end obtains the correct physical states for boson and fermions as
\begin{equation}
\begin{array}{c}
|\Psi>_{boson}=\left(1-\exp\left(-8\pi M_{E}\omega\right)\right)^{\frac{1}{2}}\sum_{n}\exp\left(-4\pi nM_{E}\omega\right)|n_{out}^{(L)}>\otimes|n_{out}^{(R)}>\\
\\
|\Psi>_{fermion}=\left(1+\exp\left(-8\pi M_{E}\omega\right)\right)^{-\frac{1}{2}}\sum_{n}\exp\left(-4\pi nM_{E}\omega\right)|n_{out}^{(L)}>\otimes|n_{out}^{(R)}>
\end{array}\label{eq: physical states-1}
\end{equation}
and the correct distributions as 
\begin{equation}
\begin{array}{c}
<n>_{boson}=\frac{1}{\exp\left(8\pi M_{E}\omega\right)-1}=\frac{1}{\exp\left[4\pi\left(2M-\omega\right)\omega\right]-1}\\
\\
<n>_{fermion}=\frac{1}{\exp\left(8\pi M_{E}\omega\right)+1}=\frac{1}{\exp\left[4\pi\left(2M-\omega\right)\omega\right]+1},
\end{array}\label{eq: final distributions}
\end{equation}
 which represent the distributions associated to the probability of
emission (\ref{eq: Corda Probability}).

Again, we emphasize that this deviation from strict thermality is
consistent with unitarity \cite{key-3,key-8,key-9} and has profound
implications for the black hole information puzzle because arguments
that information is lost during black hole's evaporation rely in part
on the assumption of strict thermal behavior of the spectrum \cite{key-3,key-8,key-9,key-17}.
In other words, the process of black hole's evaporation should be
unitary, information should be preserved and the underlying quantum
gravity theory should be unitary too. 

The main issue in this letter is the logic behind writing the expression
for the effective metric (\ref{eq: Hilbert effective}) \cite{key-18}.
We show that for an effective temperature we can write the corresponding
metric by Hawking's periodicity argument \cite{key-14,key-18,key-19,key-20}.
Let us rewrite eq. (\ref{eq: beta E}) as 
\begin{equation}
\beta_{E}(\omega)\equiv\frac{1}{T_{E}(\omega)}=\beta_{H}\left(1-\frac{\omega}{2M}\right),\label{eq: beta E-1}
\end{equation}
where $\beta_{H}\equiv\frac{1}{T_{H}}$. Following Hawking\textquoteright{}s
arguments \cite{key-19,key-20} the euclidean form of the metric will
be given by 

\begin{equation}
ds_{E}^{2}=x^{2}\left[\frac{d\tau}{4M\left(1-\frac{\omega}{2M}\right)}\right]^{2}+\left(\frac{r}{r_{E}}\right)^{2}dx^{2}+r^{2}(\sin^{2}\theta d\varphi^{2}+d\theta^{2}),\label{eq: euclidean form}
\end{equation}
which is regular at $x=0$ and $r=r_{E}$. $\tau$ is treated as an
angular variable with period $\beta_{E}(\omega)$ \cite{key-19,key-20}.
Replacing the quantity $\sum_{i}\beta_{i}\frac{\hslash^{i}}{M^{2i}}$
in \cite{key-19} with the quantity $-\frac{\omega}{2M},$ if one
follows step by step the detailed analysis in \cite{key-19} the modified
eq. (\ref{eq: Hilbert effective}) is easily obtained and one also
easily shows that $r_{E}$ in eq. (\ref{eq: euclidean form}) is the
same as in eq. (\ref{eq: effective quantities}). Thus, Hawking\textquoteright{}s
periodicity argument gives completeness to our analysis and will further
endorse the correctness of the results.

\paragraph*{Conclusion remarks}

In the remarkable paper \cite{key-7} the tunnelling approach on Hawking
radiation has been improved by explicitly finding a black body spectrum
associated with black holes. But a problem is that this result, which
has been obtained by using a reformulation of the tunnelling mechanism,
is in contrast which the other remarkable result in \cite{key-2,key-3},
that, indeed, found a probability of emission which is compatible
with a non-strictly thermal spectrum.

By using our recent introduction of an effective state for a black
hole \cite{key-8,key-9} in this paper we solved such a contradiction,
through a slight modification of the analysis in \cite{key-7}. The
final result consists in a non-strictly black body spectrum from tunnelling
mechanism.

We have also shown that, for an effective temperature, we can write
the corresponding effective metric by Hawking's periodicity arguments.
This point gave completeness to our analysis and further endorsed
the correctness of the results. 

Potential important implications for the black hole information puzzle
have been also discussed. In fact, arguments that information is lost
during black hole's evaporation partially rely on the assumption of
strict thermal behavior of the spectrum \cite{key-2,key-3,key-8,key-9}.
Hence, this letter completes our previous results in \cite{key-8,key-9}
which, by showing the black hole in terms of a well defined quantum
mechanical system, having an ordered, discrete quantum spectrum, look
consistent with the unitarity of the underlying quantum gravity theory
and with the idea that information should come out in black hole's
evaporation.

\paragraph*{Acknowledgements}

The author thanks an anonymous referee for useful criticisms and comments
which permitted to improve this letter. The author also thanks Doug
Singleton and Sujoy Modak for signalling some typos that have been
correct in this final version of the letter.

\end{document}